\documentclass[letterpaper,twocolumn,prl,showpacs,amsmath,amssymb]{revtex4}
\usepackage{mathptmx}
\usepackage[T1]{fontenc}
\usepackage{array}
\usepackage{graphicx}
\usepackage{amssymb}

\makeatletter

\providecommand{\tabularnewline}{\\}
\usepackage{hyperref}
\makeatother

\begin{document}

\title{A Magneto-Optical Trap for Polar Molecules}

\author{Benjamin K. Stuhl}
\email{stuhl@jila.colorado.edu}

\author{Brian C. Sawyer}
\author{Dajun Wang}
\author{Jun Ye}

\affiliation{JILA, National Institute of Standards and Technology and the University
of Colorado \\
Department of Physics, University of Colorado, Boulder, Colorado
80309-0440, USA}

\date{December 12, 2008}


\begin{abstract}
We propose a method for laser cooling and trapping a substantial class
of polar molecules, and in particular titanium (II) oxide (TiO). This
method uses pulsed electric fields to nonadiabatically remix the ground-state
magnetic sublevels of the molecule, allowing us to build a magneto-optical
trap (MOT) based on a quasi-cycling $J'=J''-1$ transition. Monte-Carlo
simulations of this electrostatically remixed MOT (ER-MOT) demonstrate the
feasibility of cooling TiO to a temperature of 10 $\mathrm{\mu K}$ and trapping
it with a radiation-pumping-limited lifetime on the order of 80 ms.
\end{abstract}

\maketitle

The field of ultracold polar molecules has recently made great strides.
Coherent optical transfer of magneto-associated molecules can now
produce ultracold molecular gases in the $\mathrm{X{^{1}\Sigma}}$
($v=0$) ground state with densities of $10^{12}\:\mathrm{cm^{-3}}$ and translational
temperatures of 350 nK \citep{Ni08}. Incoherent photo-association
techniques can reach the $\mathrm{X{^{1}\Sigma}}$ ($v=0$) ground
state at $100\:\mathrm{\mu K}$ \citep{Sage05}. With these temperatures
and the reasonably large electric dipole moments available from heteronuclear
bialkali molecules (e.\,g. 0.76 D for $\mathrm{X{^{1}\Sigma}}$ ($v=0$)
KRb \citep{Kotochigova03}), progress towards quantum simulations
of condensed matter systems \citep{Goral02,Micheli06} and quantum
computation \citep{DeMille02,Lee05} should be rapid. In fields such
as ultracold chemistry \citep{Hudson06}, access to molecular species
beyond the bialkali family is of great interest. Arbitrary species
can be cooled to the kelvin regime through buffer-gas cooling \citep{Weinstein98,Campbell07},
while Stark deceleration \citep{Bethlem03,Sawyer07} reaches
the tens of millikelvin level for selected light molecules. Unfortunately,
there is no demonstrated technique to further compress and cool the lukewarm
molecular clouds resulting from the latter two techniques. Even cavity-mediated
schemes for molecular laser cooling \citep{Domokos02,Andre06,Morigi07,Lev08},
while in the abstract highly attractive methods for cooling a broad,
chemically interesting set of molecules, have so far been unable to cool
these lukewarm samples, due to the schemes' low scattering rates \citep{Morigi07},
small cavity mode volumes \citep{Andre06}, and requirement
of multiparticle collective effects \citep{Domokos02,Lev08}.

Direct, free-space laser cooling and trapping would be the ideal method
for producing ultracold molecules, just as it is for atoms. Unfortunately,
atoms are in general much easier to laser cool than molecules, due to 
the latter's glaring lack of cycling transitions. Laser cooling generally requires
electronic transitions, as vibrational and rotational transitions have 
impractically long excited state lifetimes unless a cavity is used \citep{Andre06}. 
Unfortunately, these ``electronic'' transitions are never purely electronic. 
Rather, they are rovibronic, and decay into various rotational, 
vibrational, or hyperfine excited states, as well as the original ground state \citep{Bahns96}.

The branching ratios of these rovibronic decays, however, are governed
by the molecular structure and the dipole selection rules. This implies
that a clever choice of molecule can greatly reduce the number of possible
decays. Decays into excited hyperfine states are impossible in molecules with 
zero nuclear spin, as these molecules have no hyperfine structure. The Franck-Condon ratios 
for decay back to the ground vibrational level can be quite
good (99+\% for selected molecules \citep{DiRosa04}). However, the only
constraint on decays to rotationally excited levels is that all decays
satisfy the total angular momentum selection rule $\Delta J=0,\pm1$ \citep{Bethlem03}.
Thus, even without hyperfine structure, it may take up to 
three lasers per vibrational level to repump the three possible rotational decays.
However, if the ground state angular momentum $J''$ is greater than 
the excited state angular momentum $J'$, two of these three decays are forbidden.
In this case, the molecule \emph{must} follow the angular
momentum cycle $J'' \rightarrow J'=J''-1 \rightarrow J''$, and so only
one laser is required per relevant vibrational level --- making laser cooling
of these molecules truly practical.
\begin{figure}
\includegraphics{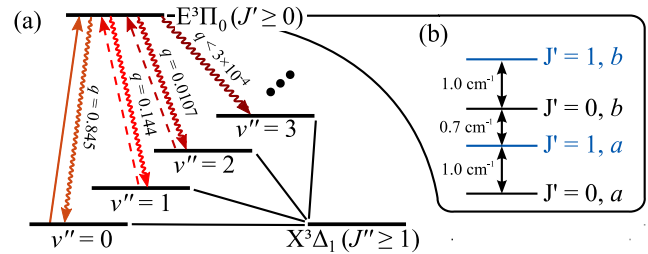}

\caption{\label{fig:level structure}(color online, not to scale). (a)~The electronic
level structure of TiO and the transitions of interest for laser cooling.
The $\mathrm{X{^{3}\Delta}}$ ground state is split by the spin-orbit
interaction into the three $\mathrm{X{^{3}\Delta_{1-3}}}$ sublevels,
of which the $\mathrm{X^{3}\Delta_{1}}$ level is the lowest.
Each sublevel contains a vibrational ladder, while each
vibrational level contains a ladder of rotationally excited states
(not shown). $\mathrm{{^{48}Ti}{^{16}O}}$ has zero nuclear spin and
thus there is no hyperfine structure. The ground-state $\Lambda$ doublet
(not shown) is much less than the natural linewidth of the $\mathrm{E{^{3}\Pi}\leftarrow X{^{3}\Delta}}$
transition. The solid arrow denotes the $v'=0\leftarrow v''=0$ P(1)-branch
cooling laser, and the dashed arrows denote the $v'=0\leftarrow v''=1$
and $v'=0\leftarrow v''=2$ P(1)-branch repump lasers. The squiggly
lines depict the dipole-allowed decays, with the associated Franck-Condon
factor $q$ \citep{Hedgecock95} next to each decay. (b)~The rotational
and $\Lambda\mathrm{-doublet}$ structure of the $\mathrm{E^{3}\Pi_{0}}$ electronic excited
state. The states are interleaved, as the rotational splitting is 
smaller than the $\Lambda\mathrm{-doublet}$ splitting; $a$ and $b$ denote the parity 
states. Both the cooling and repump lasers address the $J'=0,\:a$ state.}

\end{figure}

Thus, by combining these various transition closure criteria, we can 
identify a class of molecules that are exceptionally good candidates for laser 
cooling: they have no net nuclear spin, good Franck-Condon overlaps, and their ground 
or lowest metastable state has a higher angular momentum than the first accessible 
electronically excited state. For non-singlet molecules, the excited electronic level 
must also not be a $\Sigma$ state, as the lack of spin-orbit splitting in $\Sigma$ states means 
that the excited state can decay across the spin-orbit ladder.

We have identified a number of molecules that satisfy all of the above
requirements. TiO and TiS are both satisfactory in their absolute
ground states. Metastable FeC, ZrO, HfO, ThO, SeO, and the like are
promising \citep{DeMille08}. We expect that some other, as-yet
uncharacterized metal oxides, sulfides, and carbides should also have
the necessary electronic structure. If one is willing to accept
some hyperfine structure as the price of chemical diversity, 
some metal hydrides and metal halides may provide additional suitable candidates.

Of these candidates, we chose to focus on TiO, due to its viability
in its absolute ground state and the breadth of spectroscopy and theory
available in the literature \citep{Dobrodey01,Gustavsson91,Hedgecock95,Kobayashi02,Langhoff97,Lundevall98,Namiki98,Steimle03,Amiot95}.
A simplified level structure of TiO is shown in Fig.~\ref{fig:level structure}.
The lowest ground state of TiO is the $\mathrm{X{^{3}\Delta_{1}}}$,
with spin-orbit constant $A^{\mathrm{(X{^{3}\Delta})}}=50.61\:\mathrm{cm^{-1}}$
and rotational constant $B^{\mathrm{(X{^{3}\Delta})}}=0.534\:\mathrm{cm^{-1}}$
\citep{Amiot95}. The lowest excited state is the $\mathrm{E{^{3}\Pi_{0}}}$
with $A^{(\mathrm{E{^{3}\Pi})}}=86.82\:\mathrm{cm^{-1}}$ and $B^{(\mathrm{E{^{3}\Pi})}}=0.515\:\mathrm{cm^{-1}}$
\citep{Kobayashi02}. (While the $\mathrm{d{^{1}\Sigma^{+}}}$ and
$\mathrm{a{^{1}\Delta}}$ level are energetically below the $\mathrm{E{^{3}\Pi}}$
level, the inter-system branching ratio is expected to be very small.)
As Fig.~\ref{fig:level structure} shows, the Franck-Condon factors
\citep{Hedgecock95,Dobrodey01} for the $\mathrm{E{^{3}\Pi}-X{^{3}\Delta}}$
band are quite favorable, yielding a population leak of $3\times10^{-4}\:\mathrm{scatter^{-1}}$
(or a mean of $\sim3300$ scatters before going into a dark state)
with two repump lasers. The laser wavelengths, saturation intensities
($I_{\mathrm{sat}}$), and Franck-Condon factors for the cooling and
repump lines are summarized in Table~\ref{table:transitions}. Note
that these transitions are all accessible with diode lasers. The saturation
intensities are extremely low, as the natural linewidth $\gamma$
of the $\mathrm{E{^{3}\Pi}-X{^{3}\Delta}}$ transition is on the order
of $2\pi\times$32-40 kHz \citep{Langhoff97,Lundevall98}. This is
about 5 times weaker than the intercombination line used to build
a Yb magneto-optical trap (MOT) in \citep{Kuwamoto99}. However, with
the use of a cryogenic buffer-gas-cooled TiO source (similar to \citep{Egorov02}),
the scattering rate is still large enough to work with.

\begin{figure}
\includegraphics{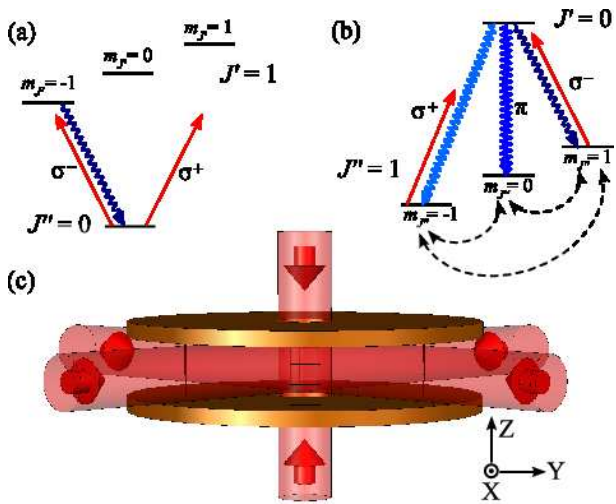}

\caption{\label{fig:ER-MOT operation}(color online). (a)~The level structure
of a traditional MOT. The local magnetic field strength and orientation
combined with the Doppler shift enhance the scattering from the laser
beam that provides the damping and restoring forces and suppresses
scattering from the counter-propagating beam. Since $J'>J''$, the
ground state(s) are always able to scatter from every beam. (b)~The
level structure of the ER-MOT. The local magnetic field still governs
which laser is preferentially scattered, but angular momentum conservation
forbids some ground states from interacting with the preferred beam.
To overcome this, the ground-state magnetic sublevel populations are
remixed by pulsed electric fields, as represented by the dashed lines.
(c)~A sample ER-MOT design. A pair of electromagnet coils are aligned
in anti-Helmholtz fashion to produce a quadrupole field. Six beams
of the cooling laser are converged on the center with their polarizations
oriented as usual for a MOT, but a set of four open-mesh grids are
added. The grids are pulsed in pairs (e.\,g., first the X-axis pair
and then the Y-axis pair) to produce the dipole electric fields needed
to remix the magnetic sublevels. The center is also illuminated
by the repump lasers (not shown).}

\end{figure}

The prospect of building a TiO MOT is tantalizing, given this quasi-closed
transition. Traditional MOTs work using a $J'=J''+1$ transition and
a magnetic field to break the degeneracy between the excited-state
magnetic sublevels. The MOT beams are polarized so that the local
orientation and strength of the quadrupole magnetic field causes the
atom to preferentially scatter from the laser beam providing a position-dependent
restoring force and a velocity-dependent damping force. The fact that
$J'>J''$ means that the atom can always scatter from the correct
beam, as shown in Fig.~\ref{fig:ER-MOT operation}(a). The standard
MOT will therefore not work for molecules using the aforementioned
$J'=J''-1$ transition {[}Fig.~\ref{fig:ER-MOT operation}(b)]. While
a magnetic field can break the degeneracy of the ground-state magnetic
sublevels and thus provide beam selectivity, the $\left|m_{J''}\right|=J''$
stretched states are effectively dark states, as they can only interact
with one of the laser beams, not both \citep{Metcalf01}.

\begin{table}
\caption{\label{table:transitions}The wavelengths, Franck-Condon factors,
and saturation intensities of the cooling and repump transitions of
TiO.}
\begin{ruledtabular}
\begin{tabular}{cc>{\centering}p{2.5cm}>{\centering}p{2.5cm}}
$\nu''$ & $\lambda_{0,\nu''}\:\mathrm{[nm]}$ & Franck-Condon factor $q_{0\nu''}$%
\footnote{from \citep{Hedgecock95}, except for the second value of $q_{00}$%
} & estimated $I_{\mathrm{sat}}\:\mathrm{[\mu W/cm^{2}]}$%
\footnote{estimated for a two-level system with $\gamma=2\pi\times32\:\mathrm{kHz}$
and scaled by $1/q_{0v''}$%
}\tabularnewline
\hline
0 & 844.7227 \citep{Kobayashi02} & 0.845; 0.872 \citep{Dobrodey01} & 8\tabularnewline
1 & 923\footnote{calculated using the diatomic molecular constants of \citep{Gustavsson91}} & 0.144 & 48\tabularnewline
2 & 1017$^{c}$ & 0.0107 & 645\tabularnewline
\end{tabular}
\end{ruledtabular}
\end{table}

What is needed, then, is a way to continually remix the ground-state
sublevels so that all the molecules spend some fraction of their time
in bright states. Fortunately, polar molecules provide just the handle
needed to accomplish this: the effective magnetic (B) and electric (E) moments
of a polar molecule depend in different ways on $m_{J}$. Thus, applying a sudden
(i.e. nonadiabatic) electric field orthogonal (or at least nonparallel)
to the local magnetic field reprojects the total angular momentum
against a new axis, randomizing $m_{J}$ (and the $\Lambda\mathrm{-doublet}$
state) by coupling the two $\Lambda\mathrm{-doublet}$ manifolds together. 
At high remix rates and high laser saturations, the molecules' time is 
equally divided across the $2\,(2\,J''+1)$ ground and $2\,J'+1$ excited states
(the factor of $2$ in the ground state is due to the electrostatic mixing of
the $\Lambda$ doublet), but they can only decay while they are in an excited state.
Thus, while the molecules are effectively always bright, the maximum photon 
scattering rate is only $\frac{2\,J'+1}{2\,(2\,J''+1)+(2\,J'+1)}\gamma$. Such
remixing of the ground-state magnetic sublevels allows the building
of a new kind of trap, the electrostatically remixed magneto-optical
trap (ER-MOT). The ER-MOT operation is shown in 
Fig.~\ref{fig:ER-MOT operation}(b). Note that, as the
local direction of the quadrupole B-field spans all of $4\pi$ steradians
over the MOT volume, a single E-field pulse will be parallel to the
local B-field in some region and therefore ineffective at remixing
the $m_{J}\mathrm{'s}$ there. This hole can, however, easily be closed
by applying a second E-field pulse, nonparallel to the first. A basic ER-MOT 
design is shown in Fig.~\ref{fig:ER-MOT operation}(c). 

To build an ER-MOT with TiO, there is a minor technical complication.
To leading order, the molecular magnetic moment can be written as
$\mu=\mu_{B}m_{J}\left(g_L\Lambda+g_S\Sigma\right)\frac{\Omega}{J\left(J+1\right)}$
Since $g_L\approx1$ and $g_S\approx2$, the magnetic moment of
the $\mathrm{X{^3\Delta_1}}$ ($\Lambda=2$ and $\Sigma=-1$) state is small, 
likely on the order of $\alpha\mu_{B}$, or the fine-structure constant times the 
Bohr magneton. In contrast, while $\Omega=0$ in the $\mathrm{E{^3\Pi_0}}$ state, the large
$\Lambda\mathrm{-doublet}$ splitting indicates strong mixing with higher
electronic excited states, and so by analogy with the $\mathrm{B{^3\Pi_0}}$
optical Zeeman measurements of \cite{Virgo05}, we estimate the 
magnetic moment to be $\sim10$ times that of the $\mathrm{X{^3\Delta_1}}$. This,
combined with the narrowness of the $\mathrm{E{^{3}\Pi}\leftrightarrow X{^{3}\Delta}}$
transition, implies that the dynamics of a TiO ER-MOT will have more in common with narrow-line
alkaline-earth MOTs \cite{Mukaiyama03} than normal alkali metal MOTs.
Given these predicted magnetic moments, the magnetic gradient in a TiO ER-MOT 
must be $\lesssim100\:\mathrm{G/cm}$. This gradient can be easily achieved
with water-cooled electromagnets \citep{Sawyer07} or rare-earth
permanent magnets \citep{Sawyer08}. In contrast, the large
($\approx3$ Debye \citep{Steimle03}) electric dipole moment of
TiO and its extremely small ground-state $\Lambda\mathrm{-doublet}$ spacing \citep{Namiki98}
mean that electric fields of only $1\:\mathrm{V/cm}$ will give Stark
shifts of about $50\gamma$ --- far more than the Zeeman shift within
the ER-MOT and thus sufficient to reproject $m_{J}$. These small
fields can easily be switched with rise times on the order of $10\:\mathrm{ns}$
(a frequency of $2800\gamma$ and 56 times the Larmor frequency due
to the electric field), and thus nonadiabaticity is assured.

A final concern regarding the viability of the TiO ER-MOT is that
either the electrostatic or magnetic fields might somehow cause
population loss by mixing in rotationally excited $J'>0$ states,
which could then decay to $J''>1$ states and be lost. Fortunately,
this loss is inhibited by the $\sim1\:\mathrm{cm^{-1}}$ rotational
splitting [Fig.~\ref{fig:level structure}(b)]. Neither the Zeeman or Stark 
shifts within the ER-MOT are anticipated to be larger than $\sim100\:\mathrm{MHz}$, and so the
perturbative probability to leave the desired $J'=0$ state is $\left(\frac{100\:\mathrm{MHz}}{30\:\mathrm{GHz}}\right)^2\approx10^{-5}$. 
This is much smaller than the $3\times10^{-4}\:\mathrm{scatter^{-1}}$ loss 
rate from decays to $v''\geq3$ and so is of no importance.

\begin{figure}
\begin{centering}
\includegraphics[scale=0.8]{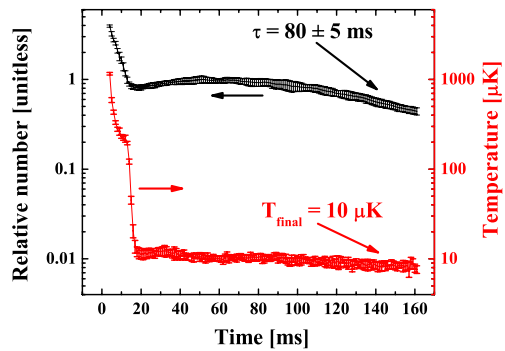}
\par\end{centering}

\caption{\label{fig:ER-MOT performance}(color online). Number (upper, black) and 
temperature (lower, red) time-of-flight 
plots for the loading of a molecular packet into a simulated TiO ER-MOT.
The initial spike on the number plot is the molecular packet flying
through the ER-MOT volume; the broad hump is the actual captured molecules.
The decay of the molecule number is due to radiation pumping of the
captured population into excited $v''\geq3$ levels, and yields an
ER-MOT lifetime of $80\pm5\:\mathrm{ms}$. Error bars represent
statistics over multiple simulation runs.}

\end{figure}

To verify the feasibility of building an ER-MOT with TiO, we performed
a set of 3-D semiclassical Monte-Carlo simulations. We conservatively
assumed a natural linewidth of $\gamma=2\pi\times32\:\mathrm{kHz}$, a 
magnetic dipole moment of $\alpha\mu_{B}$, and an electric dipole moment of 3~D.
We used a $1/e^{2}$ laser waist diameter of 6~cm. Our code treated photon 
scattering and molecule kinematics semiclassically and approximated the electrostatic 
remixing as a sudden reprojection of the diagonalized Zeeman Hamiltonian wavefunction 
against a new Stark + Zeeman Hamiltonian. In addition, the simulations used a set
of 60 additional red-detuned frequency components ($4.1\gamma$ spacing,
$7.8\:\mathrm{MHz}$ total bandwidth) within the cooling beams to
increase the capture velocity of the ER-MOT, similar to the approach
used in the Yb MOT of \citep{Kuwamoto99}. For ease of simulation,
we used a discrete molecular packet rather than a continuous source.
The packet was a 3~mm sphere initially centered at $-4.2\:\mathrm{cm}$
on the X-axis in the coordinate system of Fig.~\ref{fig:ER-MOT operation}(c).
It had a flat velocity distribution of $8\pm3.5\:\mathrm{m/s}$ centered
in orientation around $\hat{\mathrm{x}}$, with an opening half-angle of 0.3~rad.
The magnetic field gradient was $51\:\mathrm{G/cm}$ and the electric
field was pulsed to $4\:\mathrm{V/cm}$ for 100~ns at a rate of 50~kHz.
The cooling laser was detuned by $3.5\gamma$ to the red, with a saturation
parameter of $s_{\mathrm{cool}}=10.5$ per frequency component. The
repump lasers had saturations $s_{\mathrm{repump}}=6.1$. 

These simulation parameters yield the loading and temperature curves
shown in Fig.~\ref{fig:ER-MOT performance}(a) and \ref{fig:ER-MOT performance}(b),
respectively. The final temperature was approximately $10\:\mathrm{\mu K}$. The
ER-MOT lifetime was limited to about $80\:\mathrm{ms}$ by radiation pumping
into vibrationally-excited dark states. The estimated capture velocity
was $5.7\:\mathrm{m/s}$, which when taking the rotational distribution
into account allows the capture of about 0.02\% of a 4.2~K thermal
distribution. If one assumes a 4.2~K source flux of $10^{10}\:\mathrm{s^{-1}}$
\citep{Egorov02}, the calculated capture velocity, lifetime, and
ER-MOT radius predict an ER-MOT number of $10^{5}$ and a density
of $10^{9}\:\mathrm{cm^{-3}}$.

\begin{figure}
\begin{centering}
\includegraphics[scale=0.8]{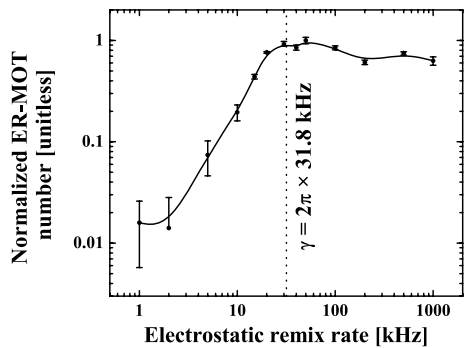}
\par\end{centering}

\caption{\label{fig:remixing action}Fractional capture vs electrostatic remix
rate after 160 ms of simulation time for the TiO ER-MOT. Error bars
represent statistics over multiple simulation runs. The curve is only
a guide to the eye.}

\end{figure}

In addition, we studied the importance of the electrostatic remixing
to the ER-MOT operation. Figure~\ref{fig:remixing action} plots the molecule
number after 160 ms against the electrostatic remix
frequency $\Gamma_{\mathrm{remix}}$. The plot clearly shows the importance
of the electrostatic remixing. At $\Gamma_{\mathrm{remix}}\ll\frac{\gamma}{2\pi}$,
no molecules are captured, since the molecules are optically pumped out
of the bright state much faster than they are remixed back into it. 
For $\Gamma_{\mathrm{remix}}\lesssim\frac{\gamma}{2\pi}$
the capture efficiency rises with increasing $\Gamma_{\mathrm{remix}}$,
and then the efficiency saturates around $\Gamma_{\mathrm{remix}}=\frac{\gamma}{2\pi}$,
as the molecules become evenly divided among the various ground sublevels.
As additional validation checks on the simulation code, we verified
that turning off the repump lasers does indeed inhibit the formation
of an ER-MOT by pumping the entire population into vibrationally excited
states. We also verified that we could reproduce the experimental
Yb MOT of \citep{Kuwamoto99} by modifying the code to simulate a
$J'=J''+1$ setup with the correct atomic parameters.

In summary, we have shown that molecules whose lowest ground- or
metastable level-intersecting electronic transition has $J'=J''-1$ 
constitute good candidates for direct laser cooling. We have found 
several molecules that satisfy this requirement and have no hyperfine 
structure. We have proposed a method to use the electric dipole moment of these 
molecules to remove dark states in the ground and thus build an 
electrostatically remixed magneto-optical trap. We have validated these ideas 
through Monte-Carlo simulation for a specific molecule, TiO, and verified 
the necessity and efficacy of the electrostatic remixing.

\begin{acknowledgments}
We thank M. Yeo,  E. Hudson, and D. DeMille for valuable discussions and
thank NIST, DOE, and NSF for support.
\end{acknowledgments}

\bibliographystyle{apsrev}

\end{document}